\RequirePackage{fix-cm}
\documentclass[twocolumn,epjc3]{svjour3}
\smartqed  
\RequirePackage{graphicx}
\journalname{Eur. Phys. J. C}
\begin{document}

\title{Teleparallel dark energy model with a fermionic field via Noether symmetry
}


\author{Yusuf Kucukakca\thanksref{e1}
}

\thankstext{e1}{e-mail: ykucukakca@akdeniz.edu.tr}


\institute{Department of Physics, Faculty of Science, Akdeniz
University, 07058 Antalya, Turkey}

\date{Received: date / Accepted: date}

\maketitle

\begin{abstract}
In the present work, we consider a model with a fermionic field that is
non-minimally coupled to gravity in the framework of teleparallel gravity.
In order to determine forms of  the coupling and potential function of fermionic field for the considered model,
we use Noether symmetry approach.
By applying this approach, for the Friedman-Robertson-Walker (FRW)
metric, we obtain respective potential and coupling function as a linear and
power-law form of the bilinear $\Psi$. Further we search the exact cosmological solution of the model.
It is shown that the fermionic field plays role of the dark energy.

\end{abstract}
\section{Introduction}
\label{intro} In the modern cosmology, it is  widely accepted that cosmic acceleration
called inflation occurred in the very early universe prior to boht
radiation and matter-dominated epoches. The idea of inflation was
originally proposed in the early 1980s  by Alan Guth to solve
several cosmological puzzles such as the flatness and horizon
puzzles \cite{guth81}. After the radiation and matter-dominated
epoches where the universe is in a decelerated expansion phase, as indicated by recent
astrophysical observations of the supernovae type Ia \cite{riess99,Perll,Sperl}
and cosmic microwave background radiation \cite{netterfield02,bennet},
another cosmic acceleration occurred in the late-time
universe. The source for this late-time acceleration was dubbed as dark
energy for which the origin has not been identified yet although several candidates are listed
in the literature. The simplest candidate
for the dark energy is the cosmological constant or the vacuum
energy. Despite its agreement with the observational data,
this model is facing serious problems of cosmological constant (see for the review papers
\cite{sahni00,carroll01,peebles03,copeland06}).

In order to resolve issue with the cosmic accelerated expansion of the universe,
two approaches have been proposed. The first approach is to take a scalar or fermionic fields as a
matter content of the universe constituting the right-hand side of the Einstein field equations.
This approach includes a variety of scalar fields
such as a quintessence \cite{Rat}, phantom \cite{Cad}, quintom \cite{Li}, tachyon \cite{SenA},
k-essence \cite{Armen}, or fermion fields \cite{Saha1,Ar,Saha2,Rib1,Rib2,Kremer,vakili2,Kremer2,rak,Liu,Kremer3,ryb}.
The second approach is to modify the geometric part of the Einstein field equations. The $f(R)$ \cite{Sot,Fel,Noj2,Cap1},
Gauss-Bonnet \cite{Gauss}, $f(R,T)$  \cite{Harr},
$f(T)$ gravities \cite{Beng,Lind,bmb,Wu,Myr} are such models belonging to second approach.

The teleparallel theory of gravity, also a teleparallel equivalent of General Relativity (GR),
was propounded by Einstein with the aim of unifying the gravity and electromagnetism \cite{Ein1,Ein2}.
Teleparallel theory is constructed by using the Weitzenbock connection,
hence its Lagrangian density is described by a
torsion scalar $T$ instead of the curvature scalar $R$ in GR that is formulated with the Levi-Civita connection.
In this theory, the dynamical variables represented by the four linearly
independent vierbein (or tetrad) fields which
play a similar role to the metric tensor in GR. The field equations of teleparallel gravity are obtained by taking
variation of the action with respect to the vierbein
fields \cite{Hay}. Recently, an interesting modified gravity by extending the teleparallel theory,
so-called $f(T)$ gravity, is proposed to explain the current accelerating expansion of the universe
without introducing the matter component \cite{Beng,Lind,bmb,Wu,Myr}. In the recent literature, to check
whether $f(T)$ gravity can be an alternative gravitational theory to the general relativity, its
various properties have been diversely investigated. We refer the reader to e.g.
\cite{a1,a2,a3,a4,a5,a6,a7,a8,a9,a10,a11,a12,a13,a14,a15,a16,a17,a18,a19,a20,a21} for some relevant works.
An other extension of teleparallel gravity can be made by introducing
a scalar field which is non-minimally coupled to the torsion scalar.
This can be regarded as a scalar-teleparallel theory of gravity, a modification of
teleparallel gravity analogous to the scalar-tensor theory as a modification of the GR. That has
recently been proposed as an alternative dark energy model \cite{geng1,geng2,Wei1,Xu1,Bani,Gu,kucuk,sad,ota1,geng3,ota2,geng4}.
The theory was called "teleparallel dark energy".
It has been found that such a theory has a richer structure than the
same one in the framework of general relativity. The richer
structure of non-minimally coupled scalar field with the torsion
scalar is due to exhibiting quintessence-like or phantom-like
behavior, or experiencing the phantom divide crossing in this
theory. We note that in the minimal coupling case, cosmological
model with the quintessence scalar field in teleparallel gravity is
identical to that in the GR.

On the other hand, some cosmological models were also investigated in
the literature by considering fermionic field (Dirac, or spinor field) as sources of the
gravitational field in the framework of GR. In this sense, to describe both
early time inflation and late-time acceleration of the universe, the models
have been proposed by using the dynamics of fermion fields with suitable interaction
potentials, where the fermion fields play the role of the inflaton
or dark energy  \cite{Saha1,Ar,Saha2,Rib1,Rib2,Kremer,vakili2,Kremer2,rak,Liu,Kremer3,ryb}. Note that in these works, the fermionic field is a classical fermion field which are presented in details in  \cite{Ar}. Recently, we have also studied
the fermionic fields as a source of inflation and dark energy in a $2 + 1$ dimensional
gravity \cite{sucu}. In addition, some cosmological solutions have
been examined with the presence of fermionic field in gravitational
theories with non-vanishing torsion \cite{carloni} and $f(T)$ gravity \cite{ratbay}.

 In the present study, motivated by the teleparallel dark energy
scenario and roles of the fermionic field in the cosmological context,
we propose a fermionic teleparallel dark energy model in which
the fermionic field with a potential non-minimal couples to the torsion scalar.
Note that the model is completely equivalent to the standard GR when the fermion field
minimally coupled to the torsion scalar \cite{Kremer}. In such a model, we need
to determine the forms of the coupling function $F(\Psi)$ and potential $V(\Psi)$.
Noether symmetry approach introduced by de Rittis et al. and Cappoziello et al., allows one to determine the
potential and the coupling function dynamically in the scalar-tensor
gravity theory \cite{rit1,rit2,cap3,cap4}. Utilizing this approach, we find the potential and
the coupling function in the teleparallel dark energy scenario with the fermionic field.
We analytically solve the field equations of the theory evolving in a spatially-flat Friedmann-Robertson-Walker spacetime.
Our results show that the fermionic teleparallel dark energy equation of state
parameter has both a quintessence, and phantom phase in this theory.

The structure of this paper is the following. In Section \ref{b1}, the
field equations are derived from a point-like Lagrangian in a
Friedman-Robertson-Walker spacetime, which is obtained from an
action including the fermionic field non-minimally coupled to the
torsion scalar in the framework of teleparallel gravity. In
Section \ref{b2}, we search the Noether symmetry of the Lagrangian
of the theory and in Section \ref{b3}, we give the exact solutions of the field
equations by using the coupling function and potential obtaining Noether symmetry approach. Finally, in the
Section \ref{conclusions}, we conclude with a brief summary of the
obtained results. It should be noted that we fully adopt the natural system of units
by taking $8\pi G = c = \hbar = 1$.  Indices $i,j,l$ run from $1$ to $4$ throughout this paper.

\section{\label{b1}
The Action and The Field Equations}
 The model considered in this work is described by the action for a
fermion field that is non-minimally coupled with the torsion scalar
\begin{eqnarray}
 \label{act1} \mathcal{A}  = \int{ d^4  x e \bigg\{F(\Psi) T
 +\frac{\imath}{2} [\bar{\psi}\Gamma^{\mu}
(\overrightarrow{ \partial_{\mu}}-\Omega_{\mu})\psi  } \nonumber \qquad \\ {
-\bar{\psi}(\overleftarrow{\partial_{\mu}}+\Omega_{\mu})
\Gamma^{\mu}\psi]-V(\Psi)}\bigg\},
\end{eqnarray}
where $e = det(e_{\mu}^a)=\sqrt{-g}$ that $e_{\mu}^a$ is tetrad
(vierbein) basis,  $T$ is a torsion scalar, $\psi$ and $\bar{\psi} = \psi^{\dag}\gamma^{0}$
denote the spinor field and its adjoint, with the dagger representing
complex conjugation. $F(\Psi)$ and
$V(\Psi)$ are generic functions, representing the coupling with gravity
and the self-interaction potential of the fermionic field
respectively. In this study, since we focus on
the effect of fermionic field in the context of teleparallel gravity,
we can neglect the contribution of the ordinary matter. We note that the
action in (\ref{act1}) with the torsion formulation of general relativity
including the fermionic field is completely equivalent to the
standard general relativity with the fermionic field where minimally couples
to the Ricci scalar. In our study, for simplicity, we assume that $F$ and
$V$ depend on only functions of the bilinear $\Psi=\bar{\psi}\psi$.
In the above action, furthermore, $\Omega_{\mu}$ is spin connection $\Omega_{\mu} = -\frac{1}{4} g_{\sigma\nu}\left[\Gamma^{\nu}_{\mu\lambda} - e_{b}^\nu \partial_{\mu} e_{\lambda}^b\right] \Gamma^{\sigma} \Gamma^{\lambda}$ with $\Gamma^{\nu}_{\mu\lambda}$ denoting the standard Levi-Civita connection and $\Gamma^{\mu}=e_{a}^\mu\gamma^{a}$. The $\gamma^{\mu}$ are Dirac matrices.

We will consider here the simplest homogeneous and isotropic
cosmological model, FRW, whose
spatially flat metric is given by

\begin{eqnarray} \label{FRW}
& & ds^2  = dt^2 -  a^2(t)\big[ dx^2 + dy^2 + dz^2\big],
\end{eqnarray}
where $a(t)$ is the scale factor of the Universe. In the teleparallel gravity, the torsion scalar
 corresponding to the FRW metric (\ref{FRW}) takes the form of
$T = -\frac{6 \dot{a}^2}{a^2}$,
where the dot represents differentiation with respect to cosmic time
$t$ (see Ref. \cite{Beng}). Considering the background  in Eq.(\ref{FRW}), it is possible
to obtain the point-like Lagrangian from action (\ref{act1})
\begin{eqnarray} \label{lag}
& & L = 6 F a \dot{a}^{2} -
\frac{\imath a^3}{2}\left(\bar{\psi}\gamma^{0}\dot{\psi}-\dot{\bar{\psi}}\gamma^{0}\psi\right)
+ a^{3} V,
\end{eqnarray}
here, because of homogeneity and isotropy of the metric it is
assumed that the spinor field depends only on time, i.e. $\psi =
\psi (t)$. The Dirac's equations for the spinor field $\psi$ and its
adjoint $\bar{\psi}$ are obtained from the point-like Lagrangian
(\ref{lag}) such that the Euler-Lagrange equations for $\psi$ and
$\bar{\psi}$ are
\begin{eqnarray}
& & \dot{\bar{\psi}} + \frac{3}{2}H \bar{\psi} -  \imath
(6 F' H^2+ V')\bar{\psi} \gamma^{0}=0, \label{dirac1}
\\& & \dot{\psi} + \frac{3}{2} H \psi + \imath
(6 F' H^2+V')\gamma^{0}\psi =0, \label{dirac2}
\end{eqnarray}
where $H = \dot{a}/a$ denotes the Hubble parameter and the prime denotes a derivative with respect to the bilinear
$\Psi$. On the other
hand, from the point-like Lagrangian (\ref{lag}) and by considering
the Dirac's equations, we find the acceleration equation from the Euler-Lagrange
equation for $a$,
\begin{equation} \label{acce}
\frac{\ddot{a}}{a}  =  -\frac{\rho_{_f} + 3 p_{_f}}{12 F}.
\end{equation}
Finally, we also have to consider the Hamiltonian constraint
equation ($E_L = 0$) associated with the Lagrangian (\ref{lag})
\begin{equation} \label{hamilton}
E_{L}  = \frac{\partial {L}}{\partial{\dot{a}}} \dot{a} +
\frac{\partial {L}}{\partial{\dot{\psi}}} \dot{\psi}
+{\dot{\bar{\psi}}} \frac{\partial {L}}{\partial{\dot{\bar{\psi}}}}
-L ,
\end{equation}
which yields Friedmann equation as follows
\begin{equation} \label{fried}
H^2  =  \frac{\rho_{_f}}{6 F}.
\end{equation}
In the acceleration and Friedmann equations, $\rho_{_f}$ and $p_{_f}$ are
the effective energy density and pressure of the fermion field,
respectively, so that they have the following forms
\begin{equation} \label{ener}
\rho_{_f} = V ,
\end{equation}
\begin{equation} \label{pressure}
p_{_f} = 4 F' H \dot{\Psi}+ (6 F' H^2 + V')\Psi - V.
\end{equation}
It is very hard to find solution for the equations (\ref{dirac1})-(\ref{fried})  since these are highly non-linear systems. In order to solve the field equations we have to determine a form for
the coupling function and the potential density of the theory. To do this, in the
following section we will use the Noether symmetry approach.

\section{\label{b2}The Noether symmetry approach}
 Symmetries play an important role in Theoretical Physics.
Specially, symmetries of the Lagrangian,  so-called a Noether symmetry, can be used to obtain the conserved quantities or constant of motions. Noether symmetry approach tells us that Lie derivative of the Lagrangian with respect to a given vector field ${\bf X}$ vanishes, i.e.
\begin{equation}
\pounds_{\bf X} L = 0. \label{noether}
\end{equation}  If condition (\ref{noether}) satisfy, then ${\bf X}$ is said to be a
symmetry for the dynamics derived from the Lagrangian $L$ and thus generates a
conserved quantity. In fact, the idea for application of the Noether symmetries as a cosmological tool is not new.
It has been introduced by de Ritis et al.
\cite{rit1,rit2} and Capozziello et al.
\cite{cap3,cap4}, in order to get solutions of field equations in the gravitational theories.
We also note that such a technique helps us to find the coupling and potential function restricting the
arbitrariness in a suitable way in the non-minimal coupled scalar-tensor theories (\cite{san2,camci,Basilakos,kuc2,muh,Kremer4,pali}).
Some cosmological solutions have been presented  both in the metric and Palatini $f(R)$ theory following the
Noether symmetry approach \cite{Cap08,Vakili08,ros,pal,hus,kuc3,sha,daraa}.
Noether symmetry approach is used to obtain exact forms of gravitational theories including $f(T)$ gravity in the literature \cite{wei2,ata,jam1,moh,hann,aslam,pal2,pal3}. On the other hand, some authors studied a cosmologic model in the framework of GR where a spinor field is non-minimally coupled with
the gravitational field via Noether symmetry approach \cite{Kremer}. They determined the coupling and potential density of the spinor field and showed that the spinor field behaves as an inflaton describing an accelerated inflationary scenario.
We will search the Noether symmetries for our model. In terms of the components of the spinor
field $\psi= (\psi_1, \psi_2, \psi_3, \psi_4)^T$ and its adjoint $\bar{\psi} =
({\psi_1}^\dagger, {\psi_2}^\dagger, -{\psi_3}^\dagger, -{\psi_4}^\dagger)$, the Lagrangian (\ref{lag})
can be rewritten as
\begin{eqnarray} \label{la2}
\hspace{-0.4cm} & & L = 6 F a \dot{a}^{2} - \frac{\imath
a^3}{2}\sum_{i=1}^4(\psi_{i}^\dagger\dot{\psi_{i}} -
\dot{\psi_{i}^\dagger}\psi_{i}) + a^{3} V.
\end{eqnarray}
Now we seek the condition in order that the Lagrangian
(\ref{la2}) would admit Noether symmetry. The configuration space of
this Lagrangian is $Q = (a, \psi_{j}, \psi_{j}^\dagger)$, whose
tangent space is $TQ =(a, \psi_{j}, \psi_{j}^\dagger, \dot{a},
\dot{\psi_{j}}, \dot{\psi_{j}^\dagger})$. The existence of Noether
symmetry given by the Eq.(\ref{noether}) implies the existence of a
vector field ${\bf X}$ such that
\begin{equation}\label{vecf}
{\bf X} = \alpha \frac{\partial}{\partial a} + \dot{\alpha}
\frac{\partial}{\partial \dot{a}} + \sum_{j=1}^{4}\left(\beta_{j}
\frac{\partial}{\partial \psi_{j}} + \dot{\beta_{j}}
\frac{\partial}{\partial \dot{\psi_{j}}} + \gamma_{j}
\frac{\partial}{\partial \psi_{j}^\dagger} + \dot{\gamma_{j}}
\frac{\partial}{\partial \dot{\psi_{j}^\dagger}}\right),
\end{equation}
where $\alpha, \beta_{j}$ and $\gamma_{j}$ are unknown functions of the variables $a,
\psi_{j}$ and $\psi_{j}^\dagger$. Hence the Noether condition
(\ref{noether}) leads to the following differential equations consisting of the coupled system of 19 equations
\begin{eqnarray}
& & \alpha + 2 a \frac{\partial \alpha}{\partial a} + \frac{F'}{F} a \sum_{i=1}^4
\left(\epsilon_{i} \beta_{i}\psi_{i}^\dagger + \epsilon_{i}
\gamma_{i}\psi_{i}\right)  =0,
\label{neq1}
\\& & F\frac{\partial{\alpha}}{\partial{\psi_{j}}} =0,\quad\quad F\frac{\partial{\alpha}}{\partial{\psi_{j}^\dagger}} =0, \label{neq2}
\\& & 3 \alpha \psi_{j} + a\beta_{j} - a \sum_{i=1}^4\left(\frac{\partial{\beta_{i}}}{\partial{\psi_{j}^\dagger}}\psi_{i}^\dagger -
\frac{\partial{\gamma_{i}}}{\partial{\psi_{j}^\dagger}}\psi_{i}\right)
= 0, \label{neq3}
\\& & 3 \alpha \psi_{j}^\dagger + a\gamma_{j} + a \sum_{i=1}^4\left(\frac{\partial{\beta_{i}}}{\partial{\psi_{j}}}\psi_{i}^\dagger -
\frac{\partial{\gamma_{i}}}{\partial{\psi_{j}}}\psi_{i}\right) = 0,
\label{neq4}
\\& & \sum_{i=1}^4 \left(\frac{\partial{\beta_{i}}}{\partial{a}}\psi_{i}^\dagger -
\frac{\partial{\gamma_{i}}}{\partial{a}}\psi_{i}\right)  =0 ,\label{neq5}
\\& & 3 \alpha V + a V'\sum_{i=1}^4 \left(\epsilon_{i} \beta_{i}\psi_{i}^\dagger + \epsilon_{i}\gamma_{i}\psi_{i}\right) = 0,
\label{neq6}
\end{eqnarray}
where $\epsilon_{i}=\left\{
\begin{array}{c}
1~~ ~~\textstyle{for}~~ i=1, 2\;\cr -1~~ \textstyle{for}~~ i=3,4\;
\end{array}
\right.. $ This system given by Eqs. (\ref{neq1})-(\ref{neq6}) are obtained by
imposing the fact that the coefficients of $\dot{a}^2,\dot{a},
\dot{\psi_{j}}, \dot{\psi_{j}^\dagger}, \dot{a}\dot{\psi_{j}}$ and
$\dot{a}\dot{\psi_{j}^\dagger}$ vanish.

One can see from equations (\ref{neq2}) that the coefficient $\alpha$ is only a function of
$a$. From the equation (\ref{neq6}) one can rewrite as follows
\begin{eqnarray} \label{new1}
\frac{3 \alpha V}{a V'} = - \sum_{i=1}^4 \left(\epsilon_{i} \beta_{i}\psi_{i}^\dagger + \epsilon_{i}\gamma_{i}\psi_{i}\right).
\end{eqnarray}
We put the equation (\ref{new1}) into (\ref{neq1}) and, recalling that $F$ and $V$ are only functions of $\Psi$,
the corresponding result is
\begin{eqnarray} \label{new2}
\frac{\alpha }{a } \frac{\partial{\alpha}}{\partial{a}} = \frac{3 F' V}{2 F V'} - \frac{1}{2} =n,
\end{eqnarray}
where $n$ is a constant. Then, we find $\alpha$ from the equation (\ref{new2})
\begin{eqnarray} \label{new3}
\alpha  = \alpha_{0} a^{n},
\end{eqnarray}
where $\alpha_{0}$ is an integration constant. Now, from the equations (\ref{neq4}), (\ref{neq5}) and (\ref{neq6}), after some algebraic calculations, one can obtain the solutions for the another symmetry generators $\beta_{j}$ and $ \gamma_{j}$ as follows
\begin{eqnarray} \label{gen}
\beta_{j}=-(\frac{3}{2} \alpha_{0}a^{n-1}+\epsilon_{j}\beta_{0})\psi_{j}, \nonumber\\
\gamma_{j}=-(\frac{3}{2} \alpha_{0}a^{n-1}-\epsilon_{j}\beta_{0})\psi_{j}^{\dagger},
\end{eqnarray}
where $\beta_{0}$ is a constant of integration. Using the above solution in the equations (\ref{new1}) and (\ref{new2}), the potential $U(\Psi)$ and the coupling function $F(\Psi)$ are obtained
\begin{eqnarray} \label{pot}
V(\Psi) = \lambda \Psi,
\end{eqnarray}
\begin{eqnarray} \label{coup}
F(\Psi) = f_{0} \Psi^{\frac{2 n + 1}{3}},
\end{eqnarray}
where $\lambda$ and $f_{0}$ are an another constant.

For $n=-1/2$, the coupling function given by the equation (\ref{coup}) becomes constant
so that our model is reduced to an action which contains fermion field that is minimally coupled
with the torsion scalar. Such a selection of Noether symmetry condition
for the potential function given by the equation (\ref{pot}) yields free Dirac spinor field with a mass term.
So that one can consider the mass term $m$ instead of $\lambda$.
In the next section we would search cosmological solutions of the field equations
using the obtained coupling functions $F(\Psi)$ and potential
$V(\Psi)$.
\section{\label{b3} Exact cosmological solutions}

In this section, we attempt to integrate our the dynamical
system given by eqs. (\ref{dirac1})-(\ref{fried}) analytically. Since the coupling and
potential functions depend on the bilinear function $\Psi$, using the Dirac's equations
(\ref{dirac1}) and (\ref{dirac2}) one gets
\begin{eqnarray}
& & \dot{\Psi} + 3 \frac{\dot{a}}{a}\Psi=0, \label{sol}
\end{eqnarray}
and integration gives
\begin{eqnarray}
& & \Psi = \frac{\Psi_0}{a^3}, \label{sol2}
\end{eqnarray}
where $\Psi_0$ is a constant of integration. We note that since the field equations can be directly integrable , it
is not necessary to calculate the constants of motion associated with the Noether symmetry. Also the constants of motion give no new constraint on the field equations. From the above solution, the acceleration and Friedmann equations become only a function of the cosmic scale factor and,
can be directly integrated as indicated in the following cases.
\subsection{Case $A$} \label{case1}
Firstly, we consider the fermion field is minimally coupled to the torsion scalar,
i.e. $n=-1/2$. This case has been studied in ref. \cite{Kremer}. Using the potential (\ref{pot}) in the Friedmann equation together with the equation (\ref{ener}),
the time evolution of the scale factor can be easily calculated and has the form
\begin{eqnarray}
& & a(t) = (\frac{3\lambda\Psi_0}{4})^{1/3} (t-c_{1})^{2/3}, \label{scale1}
\end{eqnarray}
here $c_{1}$ is an integration constant and we take $f_{0}=\frac{1}{2}$. The energy density and pressure of
the fermionic field follow (\ref{ener}) and (\ref{pressure}),
yielding
\begin{eqnarray}
& & \rho_{_f} = \frac{4}{3(t-c_{1})^2},
\qquad p_{_f}=0. \label{sol-eden}
\end{eqnarray}
Therefore, form this solutions we conclude that the fermionic field behaves as a
standard pressureless matter field.
\subsection{Case $B$} \label{case2}
Now, we consider general case where the coupling function $F(\Psi) = f_{0} \Psi^{\frac{2n+1}{3}}$.
The Friedmann equation for this case can be rewritten as
\begin{eqnarray}
& & \dot{a}^2= a_{0} a^{2 n},
\qquad a{0}= \frac{\lambda \Psi_{0}^{\frac{2(1-n)}{3}}}{6 {f_0}}. \label{fried2}
\end{eqnarray}
The general solution of the equation is
\begin{eqnarray}
& & a(t) = \left[a_{0}(n-1)(t-c_{2})\right]^{-\frac{1}{n-1}}, \label{scale2}
\end{eqnarray}
where $c_{2}$ is an another integration constant and $n\neq1$. Inserting the solution (\ref{scale2}) into the acceleration equation (\ref{acce}) together with the equation (\ref{ener}) and (\ref{pressure}), we get $\lambda=6f_{0}$. For $n=1$,
the coupling function reduced to the form $F(\Psi) = f_{0} \Psi$
so that the solution of equation (\ref{scale2}) for the cosmic scale factor can be obtained by $a(t)=c_{3}\exp{(\sqrt{a_{0}} t)}$
which stands for a de Sitter solution. Thus, this solution shows the fermionic field can be behaved as inflaton.

The deceleration parameter, which is an important quantity in the cosmology,
is defined by $q =-a \ddot{a}/\dot{a}^2$, where the positive sign
of $q$ indicates the standard decelerating models and the
negative sign corresponds to accelerating models. The $q=0$
corresponds to expansion with a constant velocity. It takes the
following form in this model
\begin{eqnarray} \label{dec1}
& &  q = -n.
\end{eqnarray}
From Eq. (\ref{dec1}) we see that the universe is accelerating for
$n>0$ and decelerating for $n<0$. We can also define the equation of state parameter for
the fermionic field by using Eqs. (\ref{acce})-(\ref{pressure}) as
$w_{f}\equiv\frac{P_{f}}{\rho_{f}}=\frac{2q - 1}{3}$. Then
it can be obtained by
\begin{eqnarray} \label{eos1}
& &  w_{f} = -\frac{2n+1}{3}.
\end{eqnarray}
where the time evolution of the energy density and pressure of
the fermion field read
\begin{eqnarray}
& & \rho_{f} = \lambda \Psi_{0} \left[a_{0}(n-1)(t-c_{2})\right]^{\frac{3}{n-1}} = -\frac{2n+1}{3} p_{f}. \label{enpres}
\end{eqnarray}

Cosmological observations denote that $w$ lies in a very narrow strip
close to $w = -1$. The case $w = -1$ corresponds to the cosmological
constant. For $w<-1$, the phantom phase is observed, and for $-1 < w
< -1/3$ the phase is described by quintessence. Thus, in the
interval $0 < n < 1$, we have the quintessence
phase. If $n > 1$, then the phantom phase occurs, where
the universe is both expanding and accelerating. Therefore, we conclude that the fermionic field
behaves as both the quintessence and phantom dark energy.

\section{Conclusions} \label{conclusions}
Teleparallel gravity is an equivalent formulation of GR in which instead of the curvature scalar $R$,
one utilizes the torsion scalar $T$ for the action. By extending the teleparallel gravity,
some authors have recently suggested the teleparallel dark
energy models to explain the cosmic acceleration of the
universe \cite{geng1,geng2,Wei1,Xu1,Bani,Gu,kucuk,sad,ota1,geng3,ota2,geng4}.
That was also our motivation in the present study where we proposed a new teleparallel dark
energy model in which a fermionic
field has a potential and is also non-minimally coupled to gravity
in the framework of teleparallel gravity.
Noether symmetry approach is useful
in obtaining physically viable choices of the coupling and
potential function of the fermionic field.
By applying this approach to the Lagrangian given by eq. (\ref{la2}), we have
obtained the explicit forms of the corresponding the coupling and potential function
as $V(\Psi) = \lambda \Psi$ and $F(\Psi) = f_{0} \Psi^{\frac{2 n + 1}{3}}$, respectively.
For the minimally coupled fermion field case which is equivalent of GR i.e for $n=-1/2$,
the cosmological solution shows that the fermionic field behaves like a standard pressureless matter field.
On the other hand, in the non-minimally coupled fermion field case, for $n=1$ we found the de Sitter solution,
whereas for the general $n$ we found the power law expansion for the cosmological scale factor (see Eq. (\ref{scale2})). We
have also presented the equation of state parameter of the fermionic field for our model.
It has been turned out that a phantom like dark energy for the
intervals $0 < n < 1$  and a quintessence
like dark energy for the interval $n>1$ occur. Thus an important consequence of this work is that the fermionic field
may be interpreted as a source of dark energy.

Finally,  in the framework of GR, it is important to emphasize that when a fermionic field is
non-minimally coupled to gravity,
the existence of Noether symmetry yields only a
cosmologically solution that describes the early-time accelerated expansion (see Ref. \cite{Kremer}).
While, in the framework of teleparallel gravity, this symmetry yields
cosmologically solutions that describe not only the early-time but also late-time accelerated expansion.

\begin{acknowledgements}
We are grateful to Dr. Yusuf Sucu and Dr. Timur Sahin for fruitful
discussions. This work was supported by Akdeniz University, Scientific Research
Projects Unit.
\end{acknowledgements}




\end{document}